\documentclass[preprint,review,12pt]{elsarticle}

\usepackage{graphics}

\usepackage{amssymb}
\usepackage{amsthm}
\usepackage{amsmath}
\usepackage{color}
\usepackage{epstopdf}
\usepackage{natbib}
\usepackage{epstopdf} 

\usepackage{url}

\journal{Ocean Engineering}

\begin{document}

\begin{frontmatter}

\title{An experimental comparison of velocities underneath focussed breaking waves}

\author[labelswin1,labelmelA]{Alberto Alberello\corref{cor1}}
\address[labelswin1]{Centre for Ocean Engineering Science and Technology, Swinburne University of Technology, Hawthorn, VIC 3122, Australia}
\address[labelmelA]{Department of Infrastructure Engineering, The University of Melbourne, Parkville, VIC 3010, Australia}

\cortext[cor1]{Corresponding author}

\ead{alberto.alberello@outlook.com}

\author[labamin2]{Amin Chabchoub}
\address[labamin2]{Department of Mechanical Engineering, Aalto University, FI-02150 Espoo, Finland}

\author[labelmelmech]{Jason P. Monty}
\address[labelmelmech]{Department of Mechanical Engineering, The University of Melbourne, Parkville, VIC 3010, Australia}

\author[labelswin1]{Filippo Nelli}

\author[labelmelA]{Jung Hoon Lee}

\author[labelmelmech]{John Elsnab}

\author[labelmelA]{Alessandro Toffoli}

\begin{abstract}

Nonlinear wave interactions affect the evolution of steep wave groups, their breaking and the associated kinematic field. Laboratory experiments are performed to investigate the effect of the underlying focussing mechanism on the shape of the breaking wave and its velocity field. In this regard, it is found that the shape of the wave spectrum plays a substantial role. Broader underlying wave spectra leads to energetic plungers at a relatively low amplitude. For narrower spectra waves break at a higher amplitudes but with a less energetic spiller. Comparison with standard engineering methods commonly used to predict the velocity underneath extreme waves shows that, under certain conditions, the measured velocity profile strongly deviates from engineering predictions. 
 
\end{abstract}


\end{frontmatter}


\section{Introduction}
\label{sec:intro}

Rogue waves threaten safety and survivability of marine structures. The mechanisms leading to the formation of such extreme waves have been investigated and probabilistic descriptions derived to provide improved design criteria (e.g. \cite{perlin2013breaking,bitner2014occurrence,toffoli2012statistics,alberello2016non}). Breaking of large waves is the most hazardous condition in terms of wave forces on marine structures \cite{faltinsen1993sea,grue2002four,kim2008nonlinear}. However, it remains elusive how the mechanism leading the formation of rogue wave affects the wave shape at the breaking and the associated kinematic field.

Measurements under deep water breaking waves have shown that wave velocities, and associated forces,  exceed those predicted by the potential flow theory in the crest region. Using Laser Doppler Anemometry (\textsc{lda}) under plungers, Easson \& Greated \cite{easson1984breaking} report velocities two times larger than those predicted by linear theory and forces fives times larger than those of an equivalent 5$^{th}$ order Stokes wave. Analogous results are reported in Kim et al. \cite{kim1992kinematics} for a spillers in random sea. Measured particle velocities in the crest region exceed those predicted using equivalent Stokes wave and linear superposition of the spectral components. Kim et al. \cite{kim1992kinematics} argue that the asymmetric shape (crest higher than the troughs with forward leaning wave front) associated to large transient waves as a result of energy focussing might affect the accuracy of the estimation of the velocity field.

Breaking waves have also been experimentally investigated  by means of Particle Image Velocimetry (\textsc{piv}), this technique, compared to \textsc{lda}, offers the advantage of obtaining fluid velocities over a plane (unlike pointwise \textsc{lda} measurements). Under plungers, Skyner \cite{skyner1996acomparison} recorded particle velocities higher than the phase speed of the waves. Observations of velocities exceeding the phase speed were also made by Perlin et al. \cite{perlin1996anexperimental}, even though the fluid flow presents a different topology compared to Skyner \cite{skyner1996acomparison}. Difference in the flow structure are most certainly related to a different underlying wave spectrum. \textsc{piv} was systematically employed by Grue et al. \cite{grue2003kinematics,grue2006experimental,grue2012orbital} to investigate breaking waves in deep water conditions. Monochromatic wave trains, unidirectional focussed wave groups and unidirectional random seas were all considered.

Grue et al. \cite{grue2003kinematics} observed that all velocity profiles could be described by an universal profile if opportune dimensionless parameters were chosen. The velocity profile beneath a wave can be approximated by a third order monochromatic Stokes wave with the same period and amplitude using the so-called Grue method \cite{grue2003kinematics}. The wavenumber $k$ and the steepness $\epsilon$ (product of the wavenumber and the linear wave amplitude $a$) are obtained numerically solving the system of equations:
\begin{equation}
\begin{cases}
 \dfrac{\omega^2}{gk} = 1 + \epsilon^2 \\ 
 k\eta_M = \epsilon + \dfrac{1}{2}\epsilon^2 + \dfrac{1}{2}\epsilon^3 \\
\end{cases}
\label{eq_grue}
\end{equation}
The radial frequency is computed linearly from the trough-to-trough wave period (i.e. $\omega=2\pi/T_{TT}$ being $T_{TT}$ the distance between the troughs around the crest). Once the solution is obtained the velocity profile has the exponential profile:
\begin{equation}
u_G=\epsilon \sqrt{\frac{g}{k}}\exp{(k\eta)}.
\label{eq_grue_u}
\end{equation}
The Grue velocity profile matches previous breaking measurements reported in e.g. \cite{kim1992kinematics,skyner1996acomparison,baldock1996laboratory}. Furthermore, the Grue method compares well with second order potential flow predictions \cite{stansberg2006kinematics,johannessen2010calculations}. The good performance of the Grue method and its relative simplicity established it as one of the method commonly accepted by industry standard to define the velocity profile under large waves \cite{stansberg2006kinematics}.  

Another method to estimate the velocity profile underneath a random wave field has been proposed by Donelan et al. \cite{donelan1992simple}. The method is based on a superposition of wave components. Unlike a traditional linear superposition that has been found to overestimate crest velocities, in the Donelan method spectral wave components (surface and velocity corrections) are iteratively added to the perturbed solution. To compute the velocity profile the required steps are as follows. First a Fourier Transform alghorith is used to compute amplitudes, $a_n$, and phases, $\varepsilon_n$, of the surface elevation. A vertical grid is defined, i.e. $z$. The successive velocity and amplitude increments are computed iteratively as
\begin{equation}
\delta u_n = a_n\omega_n \cos(\omega_n t+\varepsilon_n) \cdot \exp (k_n (z-\eta_{n-1})),
\label{eq_do1}
\end{equation}
\begin{equation}
u_n = u_{n-1}+\delta u_n,
\label{eq_do2}
\end{equation}
\begin{equation}
\delta \eta_n = a_n \cos(\omega_n+\varepsilon_n),
\label{eq_do3}
\end{equation}
\begin{equation}
\eta_n = \eta_{n-1}+\delta \eta_n.
\label{eq_do4}
\end{equation}
Finally, the velocities for grid points outside the water domain have to be set to zero. From the iterative procedure it can be deducted that for the $n^{th}$ component the mean water level is the pre-existing wavy surface and the velocities are computed over a varying $z$. The Donelan method has been found to compare well with field data \cite{donelan1992simple}.

In this paper the predictive performances of the Grue and Donelan methods are tested against laboratory measurements of the velocity profile underneath breaking rogue waves. The formation of the breaking waves in the wave flume is controlled by wave focussing techniques, e.g. \cite{longuet1974breaking,tromans1991new}. Two techniques commonly used in model tests are compared: the dispersive focussing \cite{longuet1974breaking,tromans1991new}, using different underlying JONSWAP spectra, and the Nonliner Schr{\"o}dinger equation (NLS) framework \cite{zakharov1968stability}. Whereas the velocity field under breaking waves generated by dispersive focussing has been examined in the past, it is yet uncertain how it compares to the kinematic field of breaking events generated using breathers solutions of the NLS that more realistically replicate wave evolution at sea.

The paper is structured as follows. In the next Section we describe the experimental set-up. The wave generation mechanisms are presented in Section \ref{ch:wg}. The evolution in space of the wave group and its spectral properties are shown in the following Section. Description of the wave shape, velocity profiles and comparison with enginnering methods is discussed in Section \ref{ch:piv}. Final remarks are reported in the Conclusions.

\section{Experimental set-up}
\label{ch:exp_su}

\begin{figure}[htbp]
\centerline{\includegraphics[trim={0 200 0 50},clip,width=0.5\textwidth]{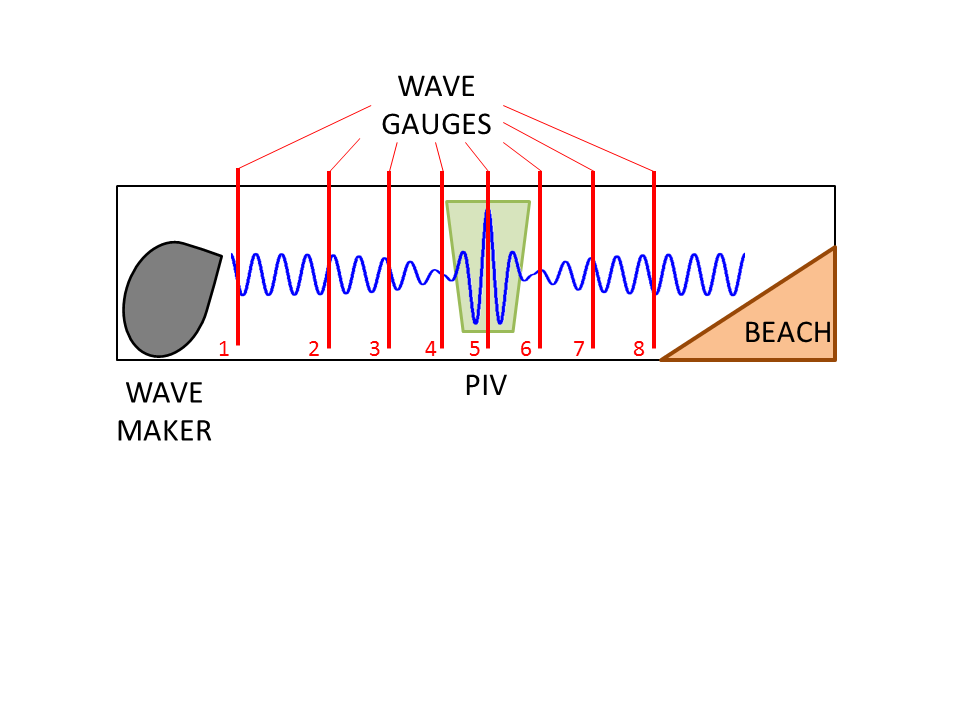}}
		\caption{Sketch of the EASI facility (not to scale).}
\label{fig:flume}
\end{figure}

The purpose of the experiments is to monitor the spatial evolution of a steep wave group and measure water particle velocity at breaking. Experiments have been conducted in the Extreme Air-Sea Interaction facility (EASI) in the Michell Hydrodynamics Laboratory at The University of Melbourne (Australia). The wave flume is 60 $\times$ 2\,m (length $\times$ width). The water depth was imposed to be 0.9\,m. At one end of the tank a computer-controlled cylindrical wave-maker produces user-defined wave forms. At the opposite end a sloping beach is installed to absorb the incoming wave energy. Optical access, to perform \textsc{piv} measurements, is provided through a glass window on the side of the flume, located 34\,m from the wave-maker. A schematic of the facility and the experimental set-up is shown in Fig.~\ref{fig:flume}.

\begin{figure}[htbp]
\centerline{\includegraphics[trim={0 120 0 80},clip,width=0.5\textwidth]{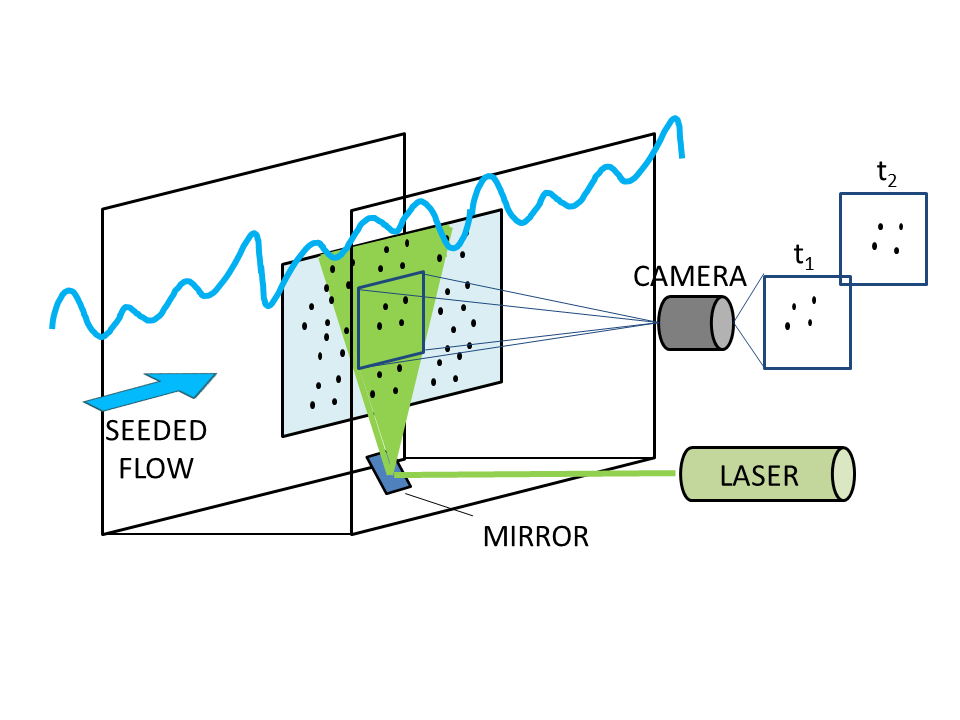}}
		\caption{Sketch of the \textsc{piv} system (not to scale).}
\label{fig:piv}
\end{figure}

At the window, the shape of the breaking wave is recorded by a camera and \textsc{piv} measurements can be undertaken. This technique has been used to explore coastal and ocean processes at laboratory scale since the 90s, e.g. \cite{greated1992particle,chang1996measurement}. \textsc{piv} allows the calculation of the spatio-temporal properties of the kinematic field by cross-correlating pairs of images of a properly seeded fluid. The analysis of two images, taken at time $\Delta t$ apart, provides the displacement of the particles and consequently their velocity \cite{adrian2011particle}. Experiments are performed with a two-dimensional \textsc{piv} set-up, i.e. only the planar velocities components along the flow and in the vertical direction are extracted. The set-up is sketched in Fig.~\ref{fig:piv}.

The laser beam is generated by Photonics DM20-527 dual head Nd:YLF laser that delivers 100\,mJ/pulse at 15\,Hz. The beam is converted in a light sheet at the centre of the tank via a series of optics. Images are recorded by Andor CMOS camera equipped with a Nikkor f/3.5 60\,mm macro lens. The camera resolution is 2120 $\times$ 2560\,pixel and the corresponding field of view is approximately 170 $\times$ 200 mm (horizontal $\times$ vertical). Silver coated glass spheres with mean particle diameter of 10 $\mu$m are used to seed the water. The laser and the particles used in the experiments provide a better image quality compared to a similar set-up used for preliminary tests \cite{alberello2016omae}. The separation time between images pairs is $\Delta \, t=2.5$\,ms. During the pre-processing step water surface is manually detected to mask the air side to improve the quality of the subsequent cross-correlation algorithm. The PIVlab tool for MATLAB \cite{thielickePIVlab,thielicke2014PIVlab} is applied to extract the velocity field in the horizontal and vertical direction.

The surface elevation is recorded by resistive wave gauges at various position along the tank. The probe positions, relative to the wave-maker, are $x \in$ 14.05, 25.15, 30.10, 32.60, 33.95, 34.90, 41.40, 45.15\,m and the beach starts at $x=51.40$\,m. The probes are not equispaced because their positioning along the tank is constrained by accessibility reasons. The fifth probe, i.e. $x=33.95$\,m, is located in the middle of the \textsc{piv} field of view to independently monitor the surface elevation where the breaking occurs. Note that during the \textsc{piv} recording the fifth probe has been removed from the camera field of view to improve the image quality for the subsequent analysis.
To obtain robust statistical results each test is repeated 20 times.

\section{Wave generation}
\label{ch:wg}

The location of the breaking event in the wave tank is controlled deterministically by means of wave focussing techniques. In deep water conditions, the dispersive focussing has been commonly used in the past. This method relies on the differential celerity of wave components of the wave spectrum, i.e. longer waves propagate faster than shorter waves. By defining an initial phase shift at the wave-maker for each spectral component, it is possible to synchronise wave components at a specific point in space and time to generate the extreme wave, e.g. \cite{skyner1996acomparison,perlin1996anexperimental,longuet1974breaking}. New Wave Theory (NWT) \cite{tromans1991new}, popular among practitioners, relies on this technique. Dispersive focussing explicitly exploits the linear properties of the waves. Corrections at higher order exist to provide a more accurate prediction, waves are in fact fully nonlinear.

Modulational instability is one of the main mechanisms leading to the growth and, eventually, breaking of rogue waves \cite{tulin1999laboratory}. This mechanism relies on the nonlinear wave-wave interactions betweneen wave components that can be accurately modelled in the framework of the Nonlinear Schr\"odinger equation (NLS). Among the large class of breather solution of the NLS, the Peregrine breather produces one rogue event starting from an almost monochromatic wave train \cite{alberello2016omae,peregrine1983water,chabchoub2011rogue}. This mechanism has been recently exploited to investigate ship response to extreme waves, e.g. \cite{onorato2013rogueplos,zhang2016modelling,klein2016peregrine}.

\subsection{Dispersive focussing}

According to the wave linear theory the surface elevation at any given time and position is provided by:
\begin{equation}
\eta (x,t) = \sum_{j=1}^N a_j \cos(\omega _jt - k_jx + \varepsilon _j).
\label{eq_nwt1}
\end{equation}
In Eqn.~(\ref{eq_nwt1}), $\eta(x,t)$ denotes the surface elevation at time $t$ and position $x$, $\omega_j$ its wave frequency, $k_j$ the wave number, $\varepsilon_j$ the phase and $a_j$ are the amplitudes of the spectral wave components. Amplitudes can be extracted from the input wave spectrum, e.g. JONSWAP. At the focussing all spectral components are in phase, we can then write:
\begin{equation}
 A=\sum_{j=1}^N a_j.
\label{eq_sum1}
\end{equation}

In our experiments amplitudes are extracted from an underlying JONSWAP spectrum. The peak wave period imposed at the wave-maker is $T_0=0.8$\,s. For the water depth 0.9\,m, this peak period guarantees deep water conditions. The corresponding wavelength is 1\,m and the associated wavenumber $k_0=2\pi$. Different peak enhancement factor $\gamma$ were analysed during the experiments, i.e. $\gamma=1,3,6$. The lower value corresponds to the Pierson-Moskowitz spectrum, $\gamma=3$ is close to the standard JONSWAP formulation, while $\gamma=6$ provides a narrower spectrum. Wave spectra are reconstructed using 256 wave components in the frequency range $0.5\leq \omega/\omega_0\leq 2$. The amplitude of each wave component is calculated as:
\begin{equation}
a_j = A\frac{S(\omega_j)}{\sum_{j=1}^N S(\omega_j)}
\label{eq_sum2}
\end{equation}
where $S(\omega)$ is the input spectrum. The input signal at the wave-maker, i.e. $\eta(x=0,t)$, is reconstructed using Eqn.~(\ref{eq_nwt1}).

The process of identifying the correct initial input surface elevation is repeated iteratively to obtain a single breaking wave at the desired location (i.e. in the camera field of view). The calibration of the breaking position is challenging \cite{tian2010energy}. Particular attention has been devoted to avoid formation of micro-breakers (or whitecapping) on the surface before the main breaking event detected with the \textsc{piv}. Two main difficulties are encountered: waves are fully nonlinear and the steepness at the breaking onset is unknown a-priori. The degree of nonlinearity is related to the wave steepness which, for the present experiments, is high. One of the main consequences is the shifting of the focussing location compared to linear theory, e.g. \cite{baldock1996extreme}. Methods have been proposed to adjust the phases \cite{chaplin1996frequency,clauss2011new,fernandez2014extreme}. However the wave shape at the breaking onset remains uncertain \cite{toffoli2010maximum} despite wave focussing experiments have shown an inverse correlation between wave steepness at the breaking onset and the spectral bandwith \cite{perlin2013breaking}.

\subsection{Nonlinear Schr{\"o}dinger Equation (NLS)}

Compared to linear potential flow theory, the NLS equation provides an enhanced description of waves nonlinear evolution. This is a solution for the slowly varying envelope and is derived from the Euler equations written in Hamiltonian form. The equation for deep water waves, first derived by Zakharov \cite{zakharov1968stability}, reads:
\begin{equation}
i \left(\frac{\partial \psi}{\partial t} + c_g \frac{\partial \psi}{\partial x}\right) - \frac{\omega_0}{8k_0^2}\frac{\partial^2 \psi}{\partial x^2} - \frac{\omega_0 k_0^2}{2}|\psi|^2\psi=0
\label{eq_NLS}
\end{equation}
where $\psi$ is the envelope, $c_g=\omega_0/(2k_0)$ the group velocity, $\omega_0$ and $k_0$ denote the wave frequency and wave number of the carrier wave component as imposed at the wave maker.
In dimensionless form, Eqn.~(\ref{eq_NLS}) becomes:
\begin{equation}
i q_\tau + q_{\chi\chi} + 2 |q|^2q=0.
\label{eq_NLSdimensionless}
\end{equation}
The transformations $\tau=-\omega_0t/(8k_0^2)$, $\chi=x-c_gt$ and $q = \sqrt{2}k_0^2\psi$ are used.
One of the exact solutions of Eqn.~(\ref{eq_NLSdimensionless}) is the Peregrine breather \cite{peregrine1983water}:
\begin{equation}
q_P (\chi,\tau) = \left( 1 - \frac{4(1+4i\tau)}{1+4\chi^2+16\tau^2}\right) e^{2i\tau}.
\label{eq_Peregrine}
\end{equation}
The surface elevation corresponding to the Peregrine breather is:
\begin{equation}
\eta_P(x,t)=\operatorname{Re}\{\psi_P\left(x,t\right)\exp\left[i\left(k_0x-\omega_0t+\varepsilon\right)\right]\} 
\end{equation}
where $\psi_P$ denotes the solution $q_P$ after transformation to dimensional variables.

Away from the focussing the surface elevation correspond to a slightly perturbed monochromatic wave train (the Peregrine solution has infinite modulation period). At the focussing, the amplification factor of the extreme wave is 3, i.e. the rogue wave is three times higher than the monochromatic wave train from which it emerges. Analogously to the dispersive focussing, the Peregrine breather leads to the formation of only one rogue event in the wave tank, i.e. the solution is doubly localised in space and time.

In the experiments, the solution is computed for a carrier wave with period $T_0=0.8$\,s. This corresponds to the peak period of the dispersive focussing cases. Similarly to the dispersive focussing case, particular attention was devoted in avoiding formations of micro-breakers before the camera field of view. Experiments under analogous conditions reported by Shemer \& Liberzon \cite{shemer2014lagrangian} suggest that a spiller can be expected in this case.

\section{Wave evolution}
\label{ch:eta}
\subsection{Dispersive focussing}

\begin{figure*}[htbp]
\centerline{\includegraphics[trim={0 0 0 0},clip,width=1\textwidth]{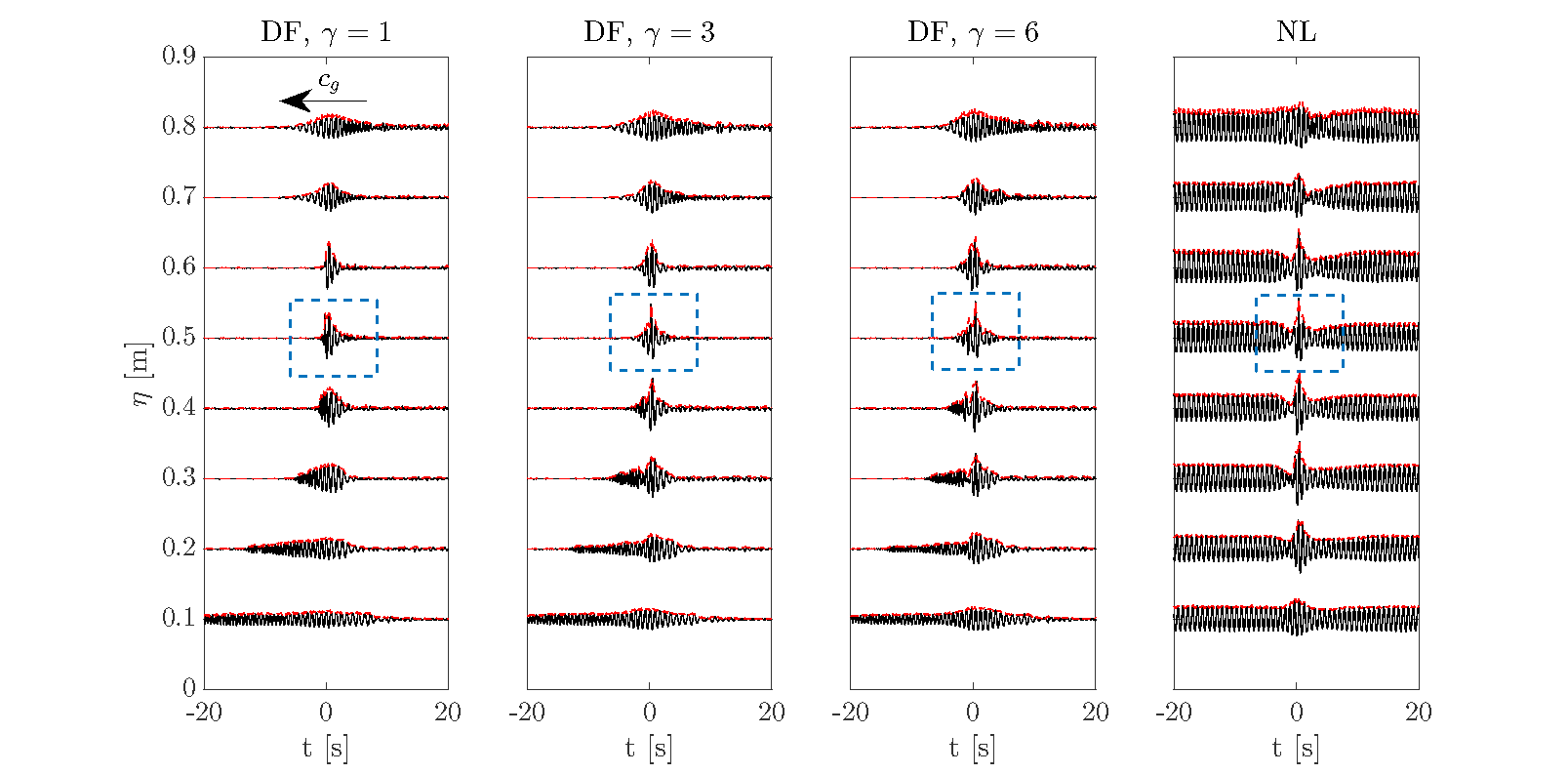}}
		\caption{Time series of the surface elevation at different distances (not equispaced) for the various wave configurations, the envelope is shown in red. Propagation is from bottom to top. The vertical shift is $0.1$\,m, an horizontal shift is applied to centre the wave-group around $t=0$\,s. The breaking location is framed. Dispersive focussing are denoted DF, Nonlinear focussing NL.}
\label{fig:tseries}
\end{figure*}

The time-series of the surface elevation are presented in Fig.~\ref{fig:tseries}. The groups become more compact in time as they approach the breaking point (probe 5). After the breaking, the wave groups broaden again, i.e. the envelope is elongated. 

\begin{figure*}[htbp]
\centerline{\includegraphics[trim={0 0 0 0},clip,width=1\textwidth]{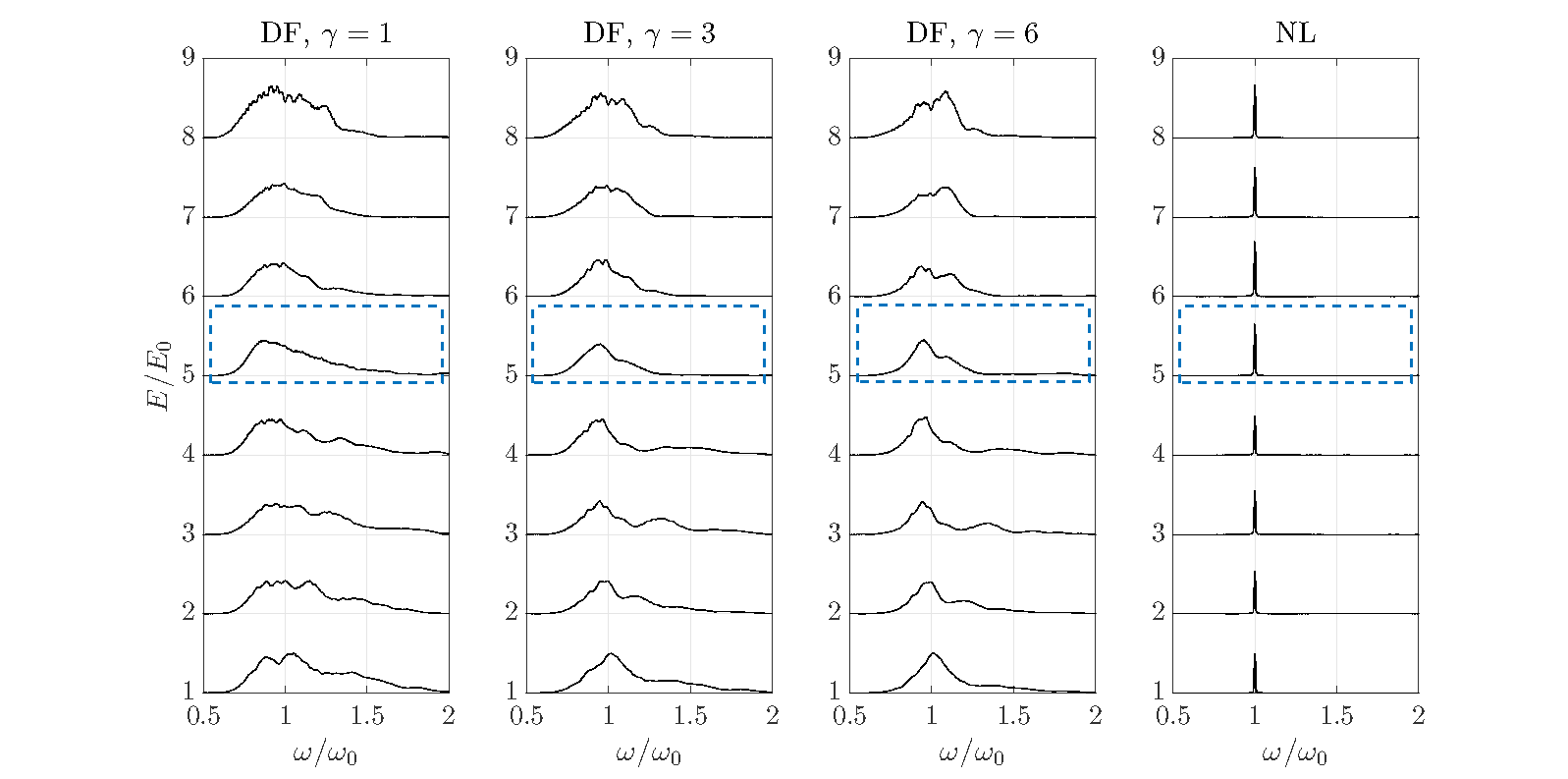}}
		\caption{Normalised surface elevation spectra at different distances (not equispaced) for the various wave configurations. Propagation is from bottom to top. The breaking location is framed.  Dispersive focussing are denoted DF, Nonlinear focussing NL.}
\label{fig:spec}
\end{figure*}

The dimensionless spectra corresponding to the various stage of evolution are reported in Fig.~\ref{fig:spec}. A spectral transformation is observed as the group propagates along the tank. An energy downshift occurs during the wave focussing (i.e. up to the breaking point). The spectral transformation is more evident for narrower spectra ($\gamma=3,6$). After the breaking energy is injected at higher frequencies ($1<\omega/\omega_0<1.5$).

Linear and secod order wave theory would not be able to predict any spectral evolution. The large steepness of the wave group at the breaking results in an high Benjamin-Feir Instability index (BFI) that underpins a strong nonlinear wave evolution. The Benjamin-Feir Instability plays a substantial role despite the fact that the dominant mechanism is dispersive focussing. Results are consistent with predictions for fully nonlinear seas \cite{socquet2005probability}.

The highly nonlinear evolution observed in the experiments also explains the difficulties encountered in calibrating the focussing location. During its evolution the group naturally tends to a more stable configuration (i.e. lower steepness) via a downshift of the energy and a consequent broadening of the spectrum itself. Due to constrains in the facility the first probe is already at about 15 wavelength from the wave maker. At this distance nonlinear evolution has already taken place. The double peaked spectral structure, particularly marked for $\gamma=1$, is consistent with fully nonlinear numerical simulations and solution of the modified NLS reported in Adcock \& Taylor \cite{adcock2016nonlinear}. Note, however, that in Adcock \& Taylor \cite{adcock2016nonlinear} breaking does not occur due to the lower steepness considered in their simulations.

To assess the quality of the focussing at the breaking location, we use a quality factor $Q$ which is the ratio between the maximum measured wave elevation and the maximum elevation of the design wave \cite{johannessen2003nonlinear}. The quality factor ranges in $0<Q<1$ with $Q=1$ corresponding to ideal focussing. In the experiments, although dispersive focussing cases break at different steepnesses, and consequently amplitudes, the input energy content at the wave maker is the same. The quality $Q$ is 0.49, 0.61 and 0.66 for $\gamma$ 1,3 and 6 respectively. By narrowing the spectrum the quality increases, i.e. the wave shape is closer to the designed one. Note that in the current experiments steeper wave conditions that lead to breaking are investigated, higher quality ($Q=0.95$) have been reported for non breaking cases \cite{johannessen2003nonlinear}.

\subsection{Nonlinear Schr{\"o}dinger Equation (NLS)}

The right panel in Fig.~\ref{fig:tseries} shows the evolution for the Peregrine solution. In this case the emergence and disappearance of the rogue wave event from an otherwise monochromatic wave train can be seen.

The spectral evolution of the Peregrine solution contrasts with the one observed for the wave groups dominated by dispersive focussing, see Fig.~\ref{fig:spec}. In this case there is no downshift of the energy. The wave nonlinear evolution is already accounted for in the equation, i.e. the NLS, but a slight broadening can be seen at the base of the spectrum (this would be clearer in logarithmic scale, cf. \cite{alberello2016omae}). Wave breaking inhibits the time-reversal symmetry \citep{chabchoub2014time} meaning that the focussing and defocussing process are asymmetric (this is clearer in the time domain, see Fig.~\ref{fig:tseries}).

A quality factor can also be defined for the Peregrine solution as the ratio between the measured amplification and the theoretical amplification, i.e. 3. In this case the quality factor is 0.9, much higher than the one recorded for the dispersive focussing cases. Using the NLS framework a wave breaking closer to the designed shape can be obtained. We must note that the Peregrine breather can be seen as a limiting case of dispersive focussing when the $\gamma$ parameter tends to infinity, i.e. extremely narrow spectrum.

\section{Wave breaking}
\label{ch:piv}
\subsection{Dispersive focussing}

\begin{figure}[htbp]
\centerline{\includegraphics[trim={0 0 0 0},clip,width=0.5\textwidth]{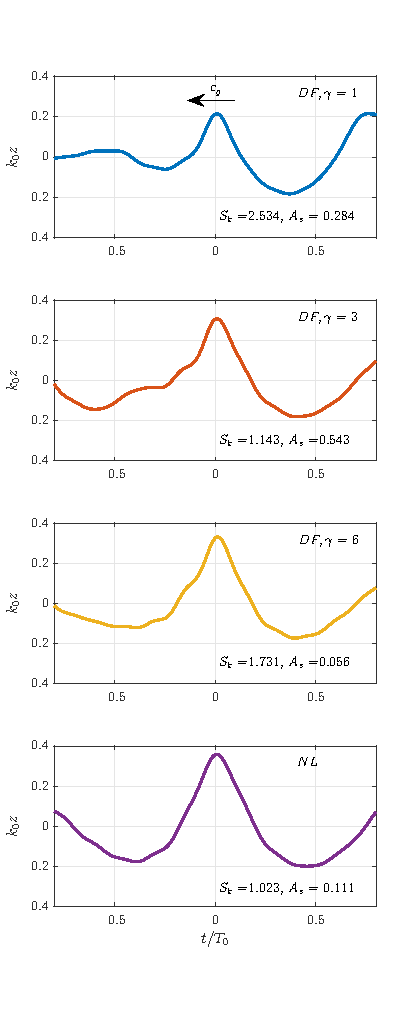}}
		\caption{Close-up of the surface elevation for the breaking wave for the various wave configurations, from top to bottom: $DF,\gamma=1$, $DF,\gamma=3$, $DF,\gamma=6$ and $NLS$. An horizontal shift is applied to obtain the crest at $t=0$\,s. Values of asymmetry are reported in the plots.}
\label{fig:tseriesz}
\end{figure}

Although the spectral evolution provides fundamental information about the nonlinear wave interactions, wave breaking is a highly localised mechanism that strongly relates to the time series rather than the spectral characteristics, e.g. \cite{chalikov2012simulation}. Fig.~\ref{fig:tseriesz} provides the dimensionless time-series at the wave breaking. Normalisation is done using the period and wavenumber imposed at the wave maker (i.e. $T_0$ and $k_0$). The asymmetry parametres are also reported and these can be used as a indication of proximity to breaking \cite{chalikov2012simulation}. $S_k$ denotes the vertical asymmetry, $A_s$ the horizontal asymmetry \cite{babanin2007predicting}. The former is a measure of how higher the crest is with respect to the trough, while the latter indicates whether the wave is leaning forward ($A_s<0$) or backwards ($A_s>0$).

In all the cases the wave period of the extreme wave is shorter than the input period, this despite the spectral downshift. Surface elevations are strongly asymmetric around the horizontal axis (i.e. $S_k$), asymmetry is less pronounced around the vertical axis (i.e. $A_s$). Most importantly, the analysis  of the timeseries shows that the breaking occurs at different steepnesses for the different cases. This further corroborates the difficulties of identifying the breaking onset a-priori; breaking most likely depends on the phase difference between spectral components and, possibly, the overall energy redestribution among wave components \cite{johannessen2001laboratory,johannessen2010calculations}. Our observations confirm that the breaking onset occurs at larger amplitudes for narrower spectra, cf. \cite{perlin2013breaking}.

\begin{figure*}[htbp]
\centerline{\includegraphics[trim={0 0 0 0},clip,width=1\textwidth]{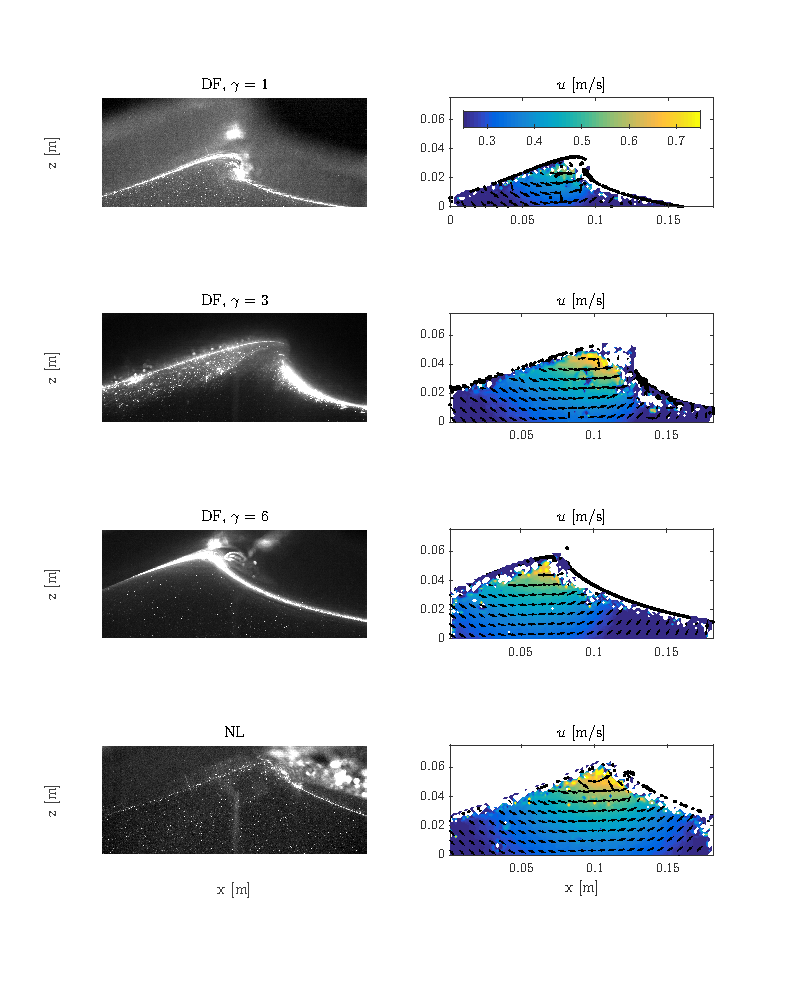}}
		\caption{\textsc{piv} images of the breaking waves (left panels) and corresponding velocity field (right panels) for the various wave configurations,from top to bottom: $DF,\gamma=1$, $DF,\gamma=3$, $DF,\gamma=6$ and $NLS$.}
\label{fig:piv_vel}
\end{figure*}

Camera images allow a detailed analysis of the shape and velocity of the breaking wave in the space domain (Fig.~\ref{fig:piv_vel}). The smallest breaking wave, the one recorded for dispersive focussing and $\gamma=1$, results in a plunger, spiller-like breaking waves are recorded for $\gamma=6$. In deep water conditions spiller and plunger are the only possible shapes of breaking waves, and the first is more frequent in the ocean \cite{duncan2001spilling}. The velocity field corresponding to the breaking wave images is shown in the right panel of Fig.~\ref{fig:piv_vel}. Larger velocities are recorded for higher wave amplitudes.

\begin{figure}[htbp]
\centerline{\includegraphics[trim={0 0 0 0},clip,width=0.5\textwidth]{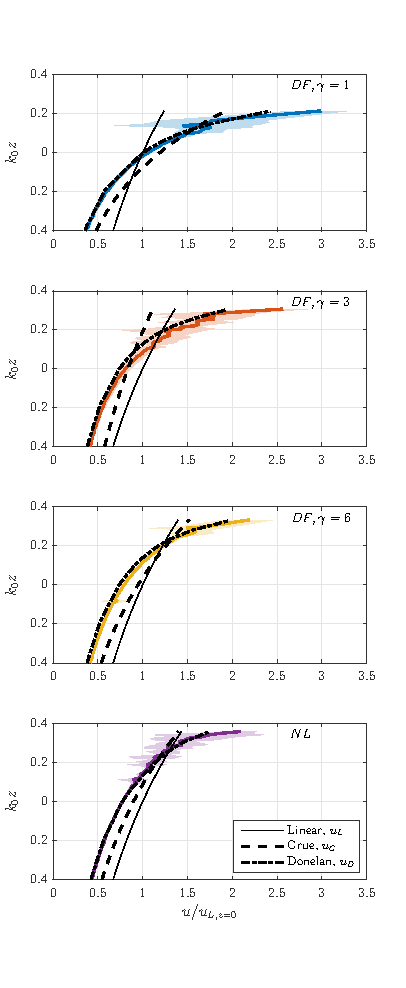}}
		\caption{Averaged horizontal velocity component for the various wave configurations, from top to bottom: $DF,\gamma=1$, $DF,\gamma=3$, $DF,\gamma=6$ and $NLS$. The confidence interval $\pm\sigma$ is shown as shaded area. The reference theoretical solutions are also shown: linear (continuous), Grue (dashed) and Donelan (dash-dotted).}
\label{fig:vel_pr_4}
\end{figure}

The dimensionless horizontal velocity profile under the crest averaged over 20 repetitions is shown in Fig.~\ref{fig:vel_pr_4}. The shaded area shows the confidence interval (i.e. $\pm$ the standard deviation $\sigma$ or  68\% confidence interval). The measured velocity is compared against the profile defined by methods commonly used in the engineering practice. The continuous line shows a reference exponential velocity profile (denoted $u_L$) of a monochromatic wave with amplitude $\eta_M$ and period $T_0$. The dashed line ($u_G$) is the profile obtained by applying the Grue method \cite{grue2003kinematics} which requires $\eta_M$ and the measured trough-to-trough period, i.e. $T_{TT}$, as inputs. The dash-dotted line ($u_D$) depicts the profile obtained by using Donelan method \cite{donelan1992simple}.

\begin{figure}[htbp]
\centerline{\includegraphics[trim={0 0 0 0},clip,width=0.5\textwidth]{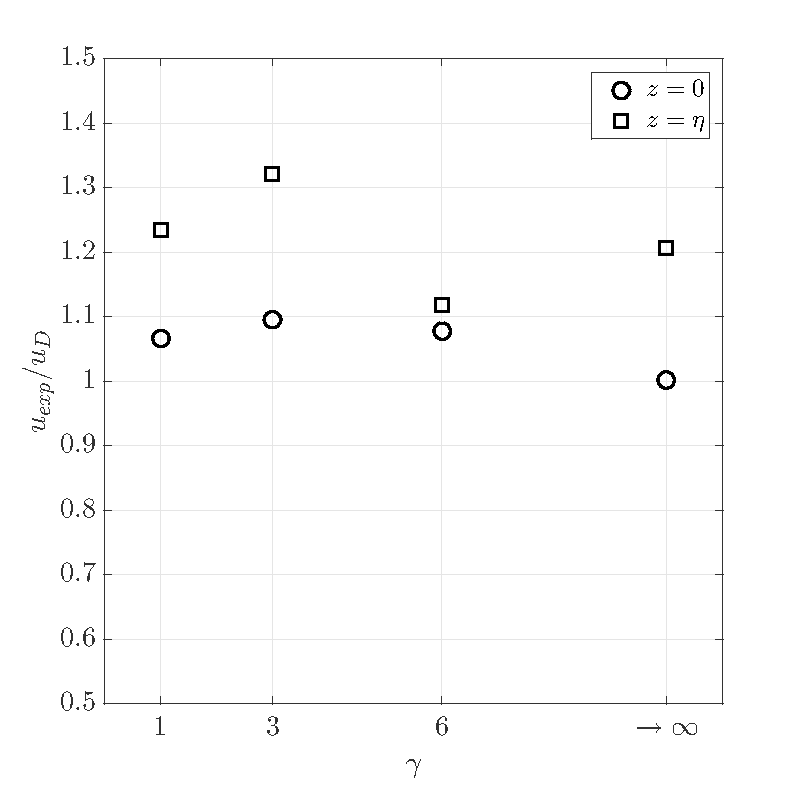}}
		\caption{Relative difference between \textsc{piv} measurements and Donelan prediction. The NLS case is denoted as $\gamma\rightarrow\infty$. Circles denote comparison of velocities extracted below the crest at the mean water lavel, squares denote comparison at the crest.}
\label{fig:vel_dif}
\end{figure}

A common feature of the \textsc{piv} experiments is the narrow confidence interval at low $z$, however, error increases close to the crest of the wave. The presence of bubbles and the highly reflective interface make the \textsc{piv} analysis challenging. As a consequence, the uncertainty in the top part of the wave is high.
Measurements show that plungers are more energetic than spillers, i.e. deviation from linear velocity profile is more accentuated for the dispersive focussing with $\gamma=1$, i.e. the dimensionless velocity $u/u_{L,z=0}$ reaches 3 at the crest ($u_{L,z=0}$ is the velocity computed from linear wave theory at $z=0$). In the dispersive focussing cases, velocities at the crest are between 2 and 3 times higher than those predicted by the linear theory. In general, the linear method leads to an overestimation of the velocity in the lower part of the wave but notably underestimates the velocity at the crest. Grue method provides a better fit but it suffers of the same drawbacks. Furthermore this method is sensitive to the definition of the local trough-to-trough wave period that might lead to underestimation of the velocities at any subsurface, see for example $\gamma=3$. Donelan method agrees better at any depth with the measured velocity. The latter uses the entire time series and not only the local properties of the breaking wave, i.e. period and amplitude. However, also the Donelan method cannot reproduce the velocity at the tip of the crest where the recorded velocities are about 25\% higher than predictions, see Fig.~\ref{fig:vel_dif}.

\subsection{Nonlinear Schr{\"o}dinger Equation (NLS)}

The solution generated using the Peregrine breather breaks at relatively high steepness, see Fig.~\ref{fig:tseriesz}. Note that the underlying spectral shape is a narrow spectrum (i.e. almost monochromatic). Narrow spectra, prone to wave growth via nonlinear mechanisms, lead to breaking waves at steepnesses closer to the Stokes limit \cite{tulin1999laboratory}. The Peregrine breather, which is driven by a nonlinear mechanism, requires an higher initial energy to evolve into a breaking shape. Compared to the dispersive focussing cases, the wave energy for the Peregrine breather is about 67\% higher. The horizontal asymmetry is rather low (i.e. a slight forward leaning is measured), the vertical asymmetry indicates that the crest is twice the trough.
 
Camera images (Fig.~\ref{fig:piv_vel}) show that the Peregrine breather breaks as a spiller under our experimental conditions, cf. \cite{shemer2014lagrangian}. In the case of the Peregrine breather, velocities at the crest are underestimated by the Grue method, whereas the Donelan method performs far better (Fig.~\ref{fig:vel_pr_4}). The presence of the breaking leads to velocities 50\% higher than those predicted by the linear model and Grue at the crest. The Donelan method reproduces the measured velocity profile at any depth. However, the Donelan method also predicts velocities at the tip of the crest are 20\% lower than those measured in the experiments, see Fig.~\ref{fig:vel_dif}.

\section{Conclusions}
\label{ch:conc}

Two different focussing techniques, dispersive focussing and Nonlinear Schr{\"o}dinger framework, are used to generate a breaking rogue wave event in a unidirectional wave flume and to compare the associated wave field. These focussing techniques are the two main mechanisms leading to the wave growth and, eventually, breaking in the ocean. The evolution has been recorded and the associated velocity field measured at the breaking by means of optical measurements, i.e. Particle Image Velocimetry (\textsc{piv}). From the experimental observations it can be inferred that the dominant generation mechanism, and hence the associated spectrum, strongly affects the wave dynamical evolution along the tank, the steepness at the breaking and the shape of the breaking wave itself. However, experiments confirm that the relation between spectral properties and time history at the breaking, in particular the breaking onset, is still elusive.

Starting from the same initial wave period, i.e. $T_0=0.8$\,s, dispersive focussing cases undergo a significant spectral change. Nonlinear interactions take place even if the main focussing mechanism is linear. The dynamical evolution is particularly evident in our experiments due to the high steepness and the long propagation distance between generation and breaking (about 35 wavelengths). Breaking occurs at higher steepness for the narrower spectrum, i.e. $\gamma=6$. However, the smaller wave is a more energetic plunger whereas for increasing $\gamma$ the breaking is gentler. Using the NLS framework to generate a breaking wave allows to account for wave nonlinear evolution. A better control on the focussing and the breaking can be achieved, i.e. the quality $Q$ is higher. In this case the steepness at the breaking is close to the Stokes limit but despite the high steepness the breaking is a less energetic spiller.

Wave velocities, measured with Particle Image Velocimetry, are compared with standard engineering models, Grue and Donelan method, that use the recorded surface elevation to derive the velocity profile. Grue method, which takes in account the maximum elevation and the local zero crossing wave period, overestimates velocities in the lower part of the wave but underestimates the highest velocities at the crest. Furthermore, Grue method is highly sensitive to the definition of the zero-crossing period; this is not straightforward in spectral conditions. Multiple wave components run over each other making the shape of the breaking wave changes rapidly in space and time domain. Donelan method, which takes in consideration the time series, provides a better fit at any elevation. However even if this more complex method is adopted the velocities in the breaking region are severely underestimated (Donelan provide velocities 20\% lower than those measured).

\section*{Acknowledgements}

A.A. and F.N. are supported by the Swinburne University of Technology Postgraduate Research Award (SUPRA). A.C. acknowledges support from the Burgundy Region, The Association of German Engineers (VDI) and the Japan Society for the Promotion of Science (JSPS).

\bibliographystyle{elsarticle-num}

\begin{thebibliography}{10}
\expandafter\ifx\csname url\endcsname\relax
  \def\url#1{\texttt{#1}}\fi
\expandafter\ifx\csname urlprefix\endcsname\relax\def\urlprefix{URL }\fi
\expandafter\ifx\csname href\endcsname\relax
  \def\href#1#2{#2} \def\path#1{#1}\fi

\bibitem{perlin2013breaking}
M.~Perlin, W.~Choi, Z.~Tian, Breaking waves in deep and intermediate waters,
  Annual Review of Fluid Mechanics 45 (2013) 115--145.
\newblock \href {http://dx.doi.org/10.1146/annurev-fluid-011212-140721}
  {\path{doi:10.1146/annurev-fluid-011212-140721}}.

\bibitem{bitner2014occurrence}
E.~Bitner-Gregersen, A.~Toffoli, Occurrence of rogue sea states and
  consequences for marine structures, Ocean Dynamics 64~(10) (2014) 1457--1468.
\newblock \href {http://dx.doi.org/10.1007/s10236-014-0753-2}
  {\path{doi:10.1007/s10236-014-0753-2}}.

\bibitem{toffoli2012statistics}
A.~Toffoli, E.~Bitner-Gregersen, S.~Suslov, M.~Onorato, Statistics of wave
  orbital velocity in deep water random directional wave fields, in: ASME 2012
  31st International Conference on Ocean, Offshore and Arctic Engineering,
  American Society of Mechanical Engineers, 2012, pp. 469--475.
\newblock \href {http://dx.doi.org/10.1115/OMAE2012-83826}
  {\path{doi:10.1115/OMAE2012-83826}}.

\bibitem{alberello2016non}
A.~Alberello, A.~Chabchoub, O.~Gramstad, A.~Babanin, A.~Toffoli,
  \href{http://www.sciencedirect.com/science/article/pii/S0378383916000028}{Non-gaussian
  properties of second-order wave orbital velocity}, Coastal Engineering 110
  (2016) 42 -- 49.
\newblock \href
  {http://dx.doi.org/http://dx.doi.org/10.1016/j.coastaleng.2016.01.001}
  {\path{doi:http://dx.doi.org/10.1016/j.coastaleng.2016.01.001}}.
\newline\urlprefix\url{http://www.sciencedirect.com/science/article/pii/S0378383916000028}

\bibitem{faltinsen1993sea}
O.~Faltinsen, Sea loads on ships and offshore structures, Vol.~1, Cambridge
  university press, 1993.

\bibitem{grue2002four}
J.~Grue, On four highly nonlinear phenomena in wave theory and marine
  hydrodynamics, Applied ocean research 24~(5) (2002) 261--274.
\newblock \href {http://dx.doi.org/10.1016/S0141-1187(03)00006-3}
  {\path{doi:10.1016/S0141-1187(03)00006-3}}.

\bibitem{kim2008nonlinear}
C.~Kim, Nonlinear waves and offshore structures, Vol.~27, World Scientific,
  2008.

\bibitem{easson1984breaking}
W.~Easson, C.~Greated,
  \href{http://www.sciencedirect.com/science/article/pii/0378383984900036}{Breaking
  wave forces and velocity fields}, Coastal Engineering 8~(3) (1984) 233 --
  241.
\newblock \href
  {http://dx.doi.org/http://dx.doi.org/10.1016/0378-3839(84)90003-6}
  {\path{doi:http://dx.doi.org/10.1016/0378-3839(84)90003-6}}.
\newline\urlprefix\url{http://www.sciencedirect.com/science/article/pii/0378383984900036}

\bibitem{kim1992kinematics}
C.~Kim, R.~Randall, S.~Boo, M.~Krafft,
  \href{http://dx.doi.org/10.1061/(ASCE)0733-950X(1992)118:2(147)}{Kinematics
  of 2‐d transient water waves using laser doppler anemometry}, Journal of
  Waterway, Port, Coastal, and Ocean Engineering 118~(2) (1992) 147--165.
\newblock \href {http://dx.doi.org/10.1061/(ASCE)0733-950X(1992)118:2(147)}
  {\path{doi:10.1061/(ASCE)0733-950X(1992)118:2(147)}}.
\newline\urlprefix\url{http://dx.doi.org/10.1061/(ASCE)0733-950X(1992)118:2(147)}

\bibitem{skyner1996acomparison}
D.~Skyner, A comparison of numerical predictions and experimental measurements
  of the internal kinematics of a deep-water plunging wave, Journal of Fluid
  Mechanics 315 (1996) 51--64.
\newblock \href {http://dx.doi.org/10.1017/S0022112096002327}
  {\path{doi:10.1017/S0022112096002327}}.

\bibitem{perlin1996anexperimental}
M.~Perlin, J.~He, L.~Bernal,
  \href{http://scitation.aip.org/content/aip/journal/pof2/8/9/10.1063/1.869021}{An
  experimental study of deep water plunging breakers}, Physics of Fluids 8~(9)
  (1996) 2365--2374.
\newblock \href {http://dx.doi.org/http://dx.doi.org/10.1063/1.869021}
  {\path{doi:http://dx.doi.org/10.1063/1.869021}}.
\newline\urlprefix\url{http://scitation.aip.org/content/aip/journal/pof2/8/9/10.1063/1.869021}

\bibitem{grue2003kinematics}
J.~Grue, D.~Clamond, M.~Huseby, A.~Jensen, Kinematics of extreme waves in deep
  water, Applied Ocean Research 25~(6) (2003) 355--366.
\newblock \href {http://dx.doi.org/10.1016/j.apor.2004.03.001}
  {\path{doi:10.1016/j.apor.2004.03.001}}.

\bibitem{grue2006experimental}
J.~Grue, A.~Jensen, Experimental velocities and accelerations in very steep
  wave events in deep water, European Journal of Mechanics-B/Fluids 25~(5)
  (2006) 554--564.
\newblock \href {http://dx.doi.org/10.1016/j.euromechflu.2006.03.006}
  {\path{doi:10.1016/j.euromechflu.2006.03.006}}.

\bibitem{grue2012orbital}
J.~Grue, A.~Jensen, \href{http://dx.doi.org/10.1029/2012JC008024}{Orbital
  velocity and breaking in steep random gravity waves}, Journal of Geophysical
  Research: Oceans 117~(C7) (2012) n/a--n/a, c07013.
\newblock \href {http://dx.doi.org/10.1029/2012JC008024}
  {\path{doi:10.1029/2012JC008024}}.
\newline\urlprefix\url{http://dx.doi.org/10.1029/2012JC008024}

\bibitem{baldock1996laboratory}
T.~Baldock, C.~Swan, P.~Taylor, A laboratory study of nonlinear surface waves
  on water, Philosophical Transactions of the Royal Society of London. Series
  A: Mathematical, Physical and Engineering Sciences 354~(1707) (1996)
  649--676.
\newblock \href {http://dx.doi.org/10.1098/rsta.1996.0022}
  {\path{doi:10.1098/rsta.1996.0022}}.

\bibitem{stansberg2006kinematics}
C.~Stansberg, O.~Gudmestad, S.~Haver, Kinematics under extreme waves, Journal
  of Offshore Mechanics and Arctic Engineering 130~(2) (2006) 021010.
\newblock \href {http://dx.doi.org/10.1115/1.2904585}
  {\path{doi:10.1115/1.2904585}}.

\bibitem{johannessen2010calculations}
T.~Johannessen, Calculations of kinematics underneath measured time histories
  of steep water waves, Applied Ocean Research 32~(4) (2010) 391--403.
\newblock \href {http://dx.doi.org/10.1016/j.apor.2010.08.002}
  {\path{doi:10.1016/j.apor.2010.08.002}}.

\bibitem{donelan1992simple}
M.~Donelan, F.~Anctil, J.~Doering, A simple method for calculating the velocity
  field beneath irregular waves, Coastal engineering 16~(4) (1992) 399--424.
\newblock \href {http://dx.doi.org/10.1016/0378-3839(92)90061-X}
  {\path{doi:10.1016/0378-3839(92)90061-X}}.

\bibitem{longuet1974breaking}
M.~Longuet-Higgins, Breaking waves in deep or shallow water, in: Proc. 10th
  Conf. on Naval Hydrodynamics, Vol. 597, 1974.

\bibitem{tromans1991new}
P.~Tromans, A.~Anaturk, P.~Hagemeijer, A new model for the kinematics of large
  ocean waves-application as a design wave, in: The First International
  Offshore and Polar Engineering Conference, International Society of Offshore
  and Polar Engineers, 1991.

\bibitem{zakharov1968stability}
V.~Zakharov, Stability of periodic waves of finite amplitude on the surface of
  a deep fluid, Journal of Applied Mechanics and Technical Physics 9~(2) (1968)
  190--194.
\newblock \href {http://dx.doi.org/10.1007/BF00913182}
  {\path{doi:10.1007/BF00913182}}.

\bibitem{greated1992particle}
C.~Greated, D.~Skyner, T.~Bruce, Particle image velocimetry (piv) in the
  coastal engineering laboratory, Coastal Engineering Proceedings 1~(23).

\bibitem{chang1996measurement}
K.~Chang, P.~Liu,
  \href{https://journals.tdl.org/icce/index.php/icce/article/view/5246}{Measurement
  of breaking waves using particle image velocimetry}, Coastal Engineering
  Proceedings 1~(25).
\newline\urlprefix\url{https://journals.tdl.org/icce/index.php/icce/article/view/5246}

\bibitem{adrian2011particle}
R.~Adrian, J.~Westerweel, Particle image velocimetry, no.~30, Cambridge
  University Press, 2011.

\bibitem{alberello2016omae}
A.~Alberello, A.~Chabchoub, A.~Babanin, J.~Monty, J.~Lee, J.~Elsnab,
  E.~Bitner-Gregersen, A.~Toffoli, The velocity field underneath linear and
  nonlinear breaking rogue waves, in: ASME 2016 35th International Conference
  on Offshore Mechanics and Arctic Engineering, American Society of Mechanical
  Engineers, 2016.

\bibitem{thielickePIVlab}
W.~Thielicke, E.~Stamhuis, Pivlab-time-resolved digital particle image
  velocimetry tool for matlab (version: 1.4)\href
  {http://dx.doi.org/10.6084/m9.figshare.1092508}
  {\path{doi:10.6084/m9.figshare.1092508}}.

\bibitem{thielicke2014PIVlab}
W.~Thielicke, E.~Stamhuis, Pivlab--towards user-friendly, affordable and
  accurate digital particle image velocimetry in matlab, Journal of Open
  Research Software 2~(1) (2014) e30.
\newblock \href {http://dx.doi.org/10.5334/jors.bl}
  {\path{doi:10.5334/jors.bl}}.

\bibitem{tulin1999laboratory}
M.~Tulin, T.~Waseda, Laboratory observations of wave group evolution, including
  breaking effects, Journal of Fluid Mechanics 378 (1999) 197--232.
\newblock \href {http://dx.doi.org/10.1017/S0022112098003255}
  {\path{doi:10.1017/S0022112098003255}}.

\bibitem{peregrine1983water}
D.~Peregrine, Water waves, nonlinear schr{\"o}dinger equations and their
  solutions, The Journal of the Australian Mathematical Society. Series B.
  Applied Mathematics 25~(01) (1983) 16--43.
\newblock \href {http://dx.doi.org/10.1017/S0334270000003891}
  {\path{doi:10.1017/S0334270000003891}}.

\bibitem{chabchoub2011rogue}
A.~Chabchoub, N.~P. Hoffmann, N.~Akhmediev,
  \href{http://link.aps.org/doi/10.1103/PhysRevLett.106.204502}{Rogue wave
  observation in a water wave tank}, Phys. Rev. Lett. 106 (2011) 204502.
\newblock \href {http://dx.doi.org/10.1103/PhysRevLett.106.204502}
  {\path{doi:10.1103/PhysRevLett.106.204502}}.
\newline\urlprefix\url{http://link.aps.org/doi/10.1103/PhysRevLett.106.204502}

\bibitem{onorato2013rogueplos}
M.~Onorato, D.~Proment, G.~Clauss, M.~Klein, Rogue waves: From nonlinear
  schr{\"o}dinger breather solutions to sea-keeping test, PloS one 8~(2) (2013)
  e54629.
\newblock \href {http://dx.doi.org/10.1371/journal.pone.0054629}
  {\path{doi:10.1371/journal.pone.0054629}}.

\bibitem{zhang2016modelling}
H.~Zhang, C.~Guedes~Soares, M.~Onorato, A.~Toffoli,
  \href{http://www.sciencedirect.com/science/article/pii/S0141118715001650}{Modelling
  of the temporal and spatial evolutions of weakly nonlinear random directional
  waves with the modified nonlinear schr{\"o}dinger equations}, Applied Ocean
  Research 55 (2016) 130 -- 140.
\newblock \href
  {http://dx.doi.org/http://dx.doi.org/10.1016/j.apor.2015.11.014}
  {\path{doi:http://dx.doi.org/10.1016/j.apor.2015.11.014}}.
\newline\urlprefix\url{http://www.sciencedirect.com/science/article/pii/S0141118715001650}

\bibitem{klein2016peregrine}
Peregrine breathers as design waves for wave-structure interaction, Ocean
  Engineering 128 (2016) 199 -- 212.
\newblock \href
  {http://dx.doi.org/http://dx.doi.org/10.1016/j.oceaneng.2016.09.042}
  {\path{doi:http://dx.doi.org/10.1016/j.oceaneng.2016.09.042}}.

\bibitem{tian2010energy}
Z.~Tian, M.~Perlin, W.~Choi, Energy dissipation in two-dimensional unsteady
  plunging breakers and an eddy viscosity model, Journal of Fluid Mechanics 655
  (2010) 217–257.
\newblock \href {http://dx.doi.org/10.1017/S0022112010000832}
  {\path{doi:10.1017/S0022112010000832}}.

\bibitem{baldock1996extreme}
T.~Baldock, C.~Swan, Extreme waves in shallow and intermediate water depths,
  Coastal Engineering 27~(1) (1996) 21--46.
\newblock \href {http://dx.doi.org/10.1016/0378-3839(95)00040-2}
  {\path{doi:10.1016/0378-3839(95)00040-2}}.

\bibitem{chaplin1996frequency}
J.~Chaplin, On frequency-focusing unidirectional waves, International Journal
  of Offshore and Polar Engineering 6~(02).

\bibitem{clauss2011new}
G.~Clauss, M.~Klein, The new year wave in a seakeeping basin: Generation,
  propagation, kinematics and dynamics, Ocean Engineering 38~(14) (2011)
  1624--1639.
\newblock \href {http://dx.doi.org/10.1016/j.oceaneng.2011.07.022}
  {\path{doi:10.1016/j.oceaneng.2011.07.022}}.

\bibitem{fernandez2014extreme}
H.~Fern{\'a}ndez, V.~Sriram, S.~Schimmels, H.~Oumeraci, Extreme wave generation
  using self correcting method—revisited, Coastal Engineering 93 (2014)
  15--31.
\newblock \href {http://dx.doi.org/10.1016/j.coastaleng.2014.07.003}
  {\path{doi:10.1016/j.coastaleng.2014.07.003}}.

\bibitem{toffoli2010maximum}
A.~Toffoli, A.~Babanin, M.~Onorato, T.~Waseda, Maximum steepness of oceanic
  waves: Field and laboratory experiments, Geophysical Research Letters 37~(5).
\newblock \href {http://dx.doi.org/10.1029/2009GL041771}
  {\path{doi:10.1029/2009GL041771}}.

\bibitem{shemer2014lagrangian}
L.~Shemer, D.~Liberzon,
  \href{http://scitation.aip.org/content/aip/journal/pof2/26/1/10.1063/1.4860235}{Lagrangian
  kinematics of steep waves up to the inception of a spilling breaker}, Physics
  of Fluids 26~(1).
\newblock \href {http://dx.doi.org/http://dx.doi.org/10.1063/1.4860235}
  {\path{doi:http://dx.doi.org/10.1063/1.4860235}}.
\newline\urlprefix\url{http://scitation.aip.org/content/aip/journal/pof2/26/1/10.1063/1.4860235}

\bibitem{socquet2005probability}
H.~Socquet-Juglard, K.~Dysthe, K.~Trulsen, H.~Krogstad, J.~Liu, Probability
  distributions of surface gravity waves during spectral changes, Journal of
  Fluid Mechanics 542 (2005) 195--216.
\newblock \href {http://dx.doi.org/S0022112005006312}
  {\path{doi:S0022112005006312}}.

\bibitem{adcock2016nonlinear}
T.~Adcock, P.~Taylor,
  \href{http://www.sciencedirect.com/science/article/pii/S0141118716301523}{Non-linear
  evolution of uni-directional focussed wave-groups on a deep water: A
  comparison of models}, Applied Ocean Research 59 (2016) 147 -- 152.
\newblock \href
  {http://dx.doi.org/http://dx.doi.org/10.1016/j.apor.2016.05.012}
  {\path{doi:http://dx.doi.org/10.1016/j.apor.2016.05.012}}.
\newline\urlprefix\url{http://www.sciencedirect.com/science/article/pii/S0141118716301523}

\bibitem{johannessen2003nonlinear}
T.~Johannessen, C.~Swan,
  \href{http://rspa.royalsocietypublishing.org/content/459/2032/1021}{On the
  nonlinear dynamics of wave groups produced by the focusing of
  surface{\textendash}water waves}, Proceedings of the Royal Society of London
  A: Mathematical, Physical and Engineering Sciences 459~(2032) (2003)
  1021--1052.
\newblock \href
  {http://arxiv.org/abs/http://rspa.royalsocietypublishing.org/content/459/2032/1021.full.pdf}
  {\path{arXiv:http://rspa.royalsocietypublishing.org/content/459/2032/1021.full.pdf}},
  \href {http://dx.doi.org/10.1098/rspa.2002.1028}
  {\path{doi:10.1098/rspa.2002.1028}}.
\newline\urlprefix\url{http://rspa.royalsocietypublishing.org/content/459/2032/1021}

\bibitem{chabchoub2014time}
A.~Chabchoub, M.~Fink,
  \href{http://link.aps.org/doi/10.1103/PhysRevLett.112.124101}{Time-reversal
  generation of rogue waves}, Phys. Rev. Lett. 112 (2014) 124101.
\newblock \href {http://dx.doi.org/10.1103/PhysRevLett.112.124101}
  {\path{doi:10.1103/PhysRevLett.112.124101}}.
\newline\urlprefix\url{http://link.aps.org/doi/10.1103/PhysRevLett.112.124101}

\bibitem{chalikov2012simulation}
D.~Chalikov, A.~Babanin, Simulation of wave breaking in one-dimensional
  spectral environment, Journal of Physical Oceanography 42~(11) (2012)
  1745--1761.
\newblock \href {http://dx.doi.org/10.1175/JPO-D-11-0128.1}
  {\path{doi:10.1175/JPO-D-11-0128.1}}.

\bibitem{babanin2007predicting}
A.~Babanin, D.~Chalikov, I.~Young, I.~Savelyev, Predicting the breaking onset
  of surface water waves, Geophysical research letters 34~(7).
\newblock \href {http://dx.doi.org/10.1029/2006GL029135}
  {\path{doi:10.1029/2006GL029135}}.

\bibitem{johannessen2001laboratory}
T.~Johannessen, C.~Swan, A laboratory study of the focusing of transient and
  directionally spread surface water waves, Proceedings of the Royal Society of
  London. Series A: Mathematical, Physical and Engineering Sciences 457~(2008)
  (2001) 971--1006.
\newblock \href {http://dx.doi.org/10.1098/rspa.2000.0702}
  {\path{doi:10.1098/rspa.2000.0702}}.

\bibitem{duncan2001spilling}
J.~Duncan, Spilling breakers, Annual Review of Fluid Mechanics 33~(1) (2001)
  519--547.
\newblock \href {http://dx.doi.org/10.1146/annurev.fluid.33.1.519}
  {\path{doi:10.1146/annurev.fluid.33.1.519}}.

\end{thebibliography}

\end{document}